# The Prevalence of Similarity of the Turbulent Velocity Profile

By David Weyburne
Air Force Research Laboratory,
2241 Avionics Circle,
WPAFB OH 45433

Castillo and George [L. Castillo and W. George, "Similarity Analysis for Turbulent Boundary Layer with Pressure Gradient: Outer Flow." AIAA J. **39**, 41 (2001)] developed a flow governing equation approach for modeling the turbulent outer boundary layer region. The approach was used to develop similarity criteria for the mean velocity and Reynolds shear stress profiles. Using the criteria as a guide, Castillo, George, and coworkers examined an extensive set of experimental datasets and claim that most of these turbulent velocity boundary layers appear to be similar boundary layers when scaled with the Zagarola and Smits [M. Zagarola and A. Smits, "Mean-flow scaling of turbulent pipe flow." J. Fluid Mech. **373**, 33 (1998)] velocity parameter. In the work herein it is shown that their success at showing scaled profile similarity in many of those datasets is flawed due to a similarity compliance problem that occurs when one combines the defect profile and the Zagarola and Smits type of velocity scaling parameter. The same problem has been identified in other papers in the literature and may in fact be widespread. We conclude that similarity of the turbulent velocity profile is not as prevalent as was claimed by Castillo, George, and coworkers. The result has implications as to the accepted paradigm of the scaling of the turbulent boundary layer.

**I. Introduction**

In a now very influential paper, Castillo and George[1] developed a flow governing equation approach to seek similarity criteria for the mean velocity and Reynolds shear stress profiles for the outer boundary layer region of the turbulent boundary layer. The development led to a less-constrained version of Clauser's[2] pressure gradient constraint parameter $\beta$. Using their new pressure gradient constraint $\Lambda$ equal to a constant as a search criterion, Castillo and George found a few turbulent boundary layer experimental datasets that exhibited profile similarity when scaled with $u_e$, the velocity at the boundary layer edge. Although this choice of velocity scaling variable satisfied all of their similarity criteria, they found that it did not produce similar profiles nearly as well as the empirical Zagarola and Smits[5] velocity scaling parameter. In fact Castillo and George,[1] Castillo and Walker,[3] and Cal and Castillo[4] (and references therein) examined an extensive set of experimental datasets and claim that most turbulent velocity boundary layers appear to be similar boundary layers when scaled with the Zagarola and Smits[5] velocity scaling parameter. This is in direct contradiction to the classical belief that similar flows for turbulent boundary layers are not common and are difficult to achieve in experiments,[2] a contradiction that Castillo and George themselves acknowledge. Castillo, George, and coworkers are not the only group to find that the Zagarola and Smits[5] velocity scaling parameter works well for finding similar behavior in turbulent boundary layer datasets. Recent turbulent boundary layer scaling reviews by Panton[6] and Buschmann and Gad-el-Hak[7] confirm the Zagarola and Smits success and thereby adding support to Castillo and George's claim that "most" turbulent boundary layer datasets display similar behavior. Although their evidence seems overwhelmingly supportive, we show that some of the supporting evidence from Castillo, George, and coworkers, as well as from the review papers is flawed. It is our contention that their success at showing profile similarity is due to an incomplete assessment of the compliance for similarity, a problem originating in the use of the defect profile together with the Zagarola and Smits type of scaling parameter to claim similar behavior.

## II. Experimental Presentation

To understand the nature of the similarity compliance problem it is necessary to start with a short review. Recall that similarity of the velocity profile for 2-D wall-bounded flows is defined as the case where two velocity profiles taken from different stations along the flow differ only by simple scaling parameters in $y$ and $u(x,y)$, where $y$ is the normal direction to the wall, $x$ is the flow direction, and $u(x,y)$ is the velocity parallel to the wall in the flow direction.[8] Rather than whole profile similarity, Castillo, George, and coworkers restricted their search to outer region similarity. For the outer region, they investigated similarity using the traditional method of plotting the defect profile defined as $u_e - u(x,y)$ where $u_e$ is the velocity at the boundary layer edge. If the profiles plot on top of one another then similarity is assumed. We have found that in all of the datasets we have cross-checked from their papers, the defect profiles scaled with the Zagarola and Smits[5] velocity scale do indeed appear to show similar behavior. However, using the same data and the same scaling parameters but replotted as scaled velocity profiles, these same datasets no longer show similar behavior. This discrepancy triggered an investigation into this paradoxical behavior.

The implications of having defect similarity but not velocity profile similarity present in a dataset will become apparent below. Before we get to that, consider three cases which highlight this behavior. In Castillo and George,[1] the authors use the $y$ length scale $\delta_{99}$ and the Zagarola and Smits[5] velocity scale $u_{ZS} = u_e \delta_1 / \delta_{99}$, where $\delta_1$ is the displacement thickness, and claim these scalings result in similarity collapse of the profile data for Clauser's[2] mild APG case. We reproduce part of their Fig. 8a here as our Fig. 1a. In Fig. 1b we plot the same velocity data using the $y$-axis scale $u/u_{ZS}$ instead of $(u_e - u)/u_{ZS}$. It is evident from Fig. 1b that the eight Clauser experimental profiles scaled with $\delta_{99}$ and $u_e \delta_1 / \delta_{99}$ do not result in similarity-like behavior in any part of the boundary layer, especially not in the outer region.

Consider a second example given by Castillo and Walker[3] in which they claim that $\delta_{95}$ and the velocity scale $u_{ZS} = u_e \delta_1 / \delta_{95}$ results in similarity collapse of the data for some of Österlund's[9] ZPG datasets. We reproduce their Fig. 3 here as Fig. 2a. In Fig. 2b we plot the same velocity data using the $y$-axis scale $u/u_{ZS}$ instead of $(u_e - u)/u_{ZS}$. Again, while the scaled defect profiles appear to be similar, the scaled velocity profile plots do not.

Finally, consider a third example. According to Cal and Castillo[4] even nonequilibrium flow cases ($\Lambda \neq$constant) can show good collapse of the defect profile when scaled as $(u_e - u)/u_{ZS}$. In Fig. 3a we reproduce part of Cal and Castillo[4] Fig. 2b. This figure shows data from Schubauer and Klebanoff[10] consisting of ten nonequilibrium profiles. The profiles shown in Fig. 3a collapse reasonably well in the outer region. Now consider the same velocity data plotted in Fig. 3b using the *y*-axis scale $u/u_{ZS}$ instead of $(u_e - u)/u_{ZS}$. It is very apparent that the ten scaled profiles in Fig. 3b do not show similar behavior in either the inner region or the outer region.

**III. Defect and Velocity Profile Similarity**

The results above point to the fact that at least for the datasets in question, defect profile similarity is present but velocity profile similarity is not present in the outer region of the boundary layer. To understand why this is important it is necessary to consider the traditional definition of similarity of the velocity profile given by Schlichting[8] together with a simple similarity equivalency derivation used by Weyburne.[12] Take the length scaling parameter as $\delta_s$ and velocity scaling parameter as $u_s$. These scaling parameters can vary with the flow direction (*x*-direction) but not in the direction perpendicular to the wall (*y*-direction). According to Schlichting,[8] the scaled velocity profile at a station $x_1$ along the wall will be similar to the scaled profile at $x_2$ if

$$\frac{u(x_1, y/\delta_s)}{u_s(x_1)} = \frac{u(x_2, y/\delta_s)}{u_s(x_2)} \quad \text{for all y.} \tag{1}$$

Using the above notations, defect profile similarity would therefore be given by

$$\frac{u_e(x_1) - u(x_1, y/\delta_s)}{u_s(x_1)} = \frac{u_e(x_2) - u(x_2, y/\delta_s)}{u_s(x_2)} \quad \text{for all y.} \tag{2}$$

By inspection, one can see that the defect profile similarity will be equivalent to velocity profile similarity if

$$\frac{u_e(x_1)}{u_s(x_1)} = \frac{u_e(x_2)}{u_s(x_2)} \quad . \tag{3}$$

Eq. 3 is an important equation. Not only is it a requirement for equivalence of the defect and velocity profile similarity but this equation also shows up as a criterion for defect similarity in flow governing equation approaches to similarity developed by Rotta[13] (see his Eq. 14.3), Townsend[14] (see his Eq. 7.2.3), Castillo and George[1] (see their Eq. 9), and Jones, Nickels, and Marusic[15] (see their Eq. 3.9 a2+a4). Note that in Rotta's formulation, Rotta assumed $u_s(x) = u_\tau(x)$ but regardless, all four formulations have the identical criterion that $u_e/u_s$ must be a constant. Furthermore, in Appendix A we offer three additional theoretical derivations that indicate that $u_e/u_s$ = constant is required for similar behavior. This point is worth emphasizing; similarity requires that $u_e/u_s$ = constant for the profiles to be similar which in turn requires that defect profile similarity must be accompanied by velocity profile similarity. You cannot have one without the other. Hence the fact that Figs. 1-3 did not show defect and velocity profile similarity simultaneously is significant and at first paradoxical.

It may tempting to try to dismiss the similarity equivalence argument offered above claiming the whole profile similarity as defined by Schlichting[8] is too restrictive requiring similarity in both the inner and outer regions. Castillo and George's[1] flow governing equation development was intentionally limited to the outer region of the boundary layer even though Castillo and George never defined exactly where the outer region boundary ends and the inner region begins. In the same spirit; examination of Eqs. 1-2 indicates that the only part of the equivalence argument that changes by restricting the argument to just the outer region is that instead of applying "for all y" in the general case it becomes "for all y in the outer region" for the case restricted to the outer boundary layer region. The requirement that $u_e / u_s$ = constant still applies based on Castillo and George's[1] own development for the outer region (see their Eq. 9) which means that defect profile similarity must be accompanied by velocity profile similarity even in the case where only the outer region is considered.

Now, if one examines the three papers by Castillo, George, and coworkers,[1,3,4] some of their plots use $u_s(x) = u_e(x)$. Inspection of Eq. 3 indicates that for this choice of the velocity scaling parameter the defect profile similarity is indeed equivalent to velocity profile similarity. However, Castillo, George, and coworkers claim that most turbulent boundary layer datasets display similarity involved the Zagarola and Smits[5] velocity scale. Substituting this velocity scale for $u_s$ in Eq. 3, then for this choice the two definitions will be equivalent if

$$\frac{\delta_{99}(x_1)}{\delta_1(x_1)} = \frac{\delta_{99}(x_2)}{\delta_1(x_2)} \quad . \tag{4}$$

Clauser[2] and Zagarola and Smits[5] point out that in general this thickness ratio behaves as

$$\frac{\delta_{99}(x)}{\delta_1(x)} \propto \sqrt{c_f(x)} \quad . \tag{5}$$

The skin friction coefficient $c_f$ is known, in general, to vary with *x.* Hence the thickness ratio changes from station to station. Therefore, one cannot assume that plots showing defect similarity using $u_{ZS}$ will necessarily also show velocity profile similarity, one must verify it. In a similar vein, one cannot assume Eq. 3 is satisfied as a similarity constraint, one must verify it. To do this one can either generate plots of the $u_e / u_s$ and show that they are constant for the profiles in question or an easier option is to show that defect profile similarity and velocity profile similarity are present simultaneously.

### IV. Different Forms of Defect Similarity

It is understandable why Castillo, George, and coworkers[1,3,4] did not offer proof that Eq. 3 was satisfied for the datasets they examined. Their search for similarity starts by looking for datasets that have the pressure gradient parameter $\Lambda$ = constant. The identified datasets are then simply plotted as defect profiles and checked for similarity. If the plots show similar behavior then the thinking was that there is no need to do any additional checking. However, from the above side-by-side plot comparisons (Figs. 1-3) which show defect profile similarity but not velocity profile similarity, it is clear that in those cases that Eq. 3 is not satisfied. This points to a logical inconsistency: defect similarity is apparently present but one of the requirements for defect similarity is not satisfied. To understand how this can happen we observe that although Eq. 2 is the formal definition of defect similarity, it is often written in the form

$$\frac{u_e(x) - u(x, y/\delta_s)}{u_s(x)} = f(y/\delta_s) \quad , \tag{6}$$

where *f* is some universal profile function of the dimensionless height. Now examination of the left-hand side of Eq. 6 reveals that there are two ways one can obtain similar behavior. If both ratios are constant on the left-hand side then the difference will be constant. A second way to make the left side invariant in *x* is to have each ratio change with *x* in such a way that the difference is constant. The plots in Figs. 1a-3a show that the left side of Eq. 6 is indeed showing similar behavior of the defect profile in the outer region. At the same time Figs. 1b-3b shows that the quantity $u(x, y/\delta_s)/u_s(x)$ is not a constant in the outer region from profile to profile. Furthermore, comparing the plots side-by-side indicates the ratio $u_e/u_s$ is also not a constant from profile to profile. Hence it must be the case that the while each ratio on the left side of Eq. 6 is changing with *x*, the difference is not. This means that the velocity ratio $u_e/u_s$ is compensating for the changes in the scaled experimental profile to produce apparent defect profile similarity. Therefore the Zagarola and Smits[5] velocity scale $u_{ZS} = u_e\delta_1/\delta_{99}$ is not removing the effects of both the upstream conditions and finite local Reynolds number on the outer velocity profile as suggested by Castillo and George,[1] it is in fact compensating for changes in the scaled velocity profiles. This compensation makes profiles that are plotted as defect profiles appear to behave similarly when in fact they are not similar according to the requirement that the ratio $u_e/u_s$ must be a constant.

The take away is that defect profile similarity can come about in two forms. True similarity occurs when both ratios on the left side of Eq. 6 are constant. We use the term "true similarity" in the sense that in this scenario, all the relevant constraints for velocity profile similarity according to the flow governing equation approach to similarity[1,13-15] have been satisfied for the outer region of the turbulent boundary layer. The second form of defect similarity we will term "false similarity" in the sense that although defect profile similarity is present, not all of the similarity criteria have been satisfied. In particular, the constraint $u_e/u_s$ equal to a constant is not satisfied. To ensure one does not have this false form of defect profile similarity one can either generate plots of the $u_e/u_s$ and show that it is constant for the profiles in question or an easier option is to generate plots of the defect profiles and plots of the velocity profile and verify that similarity is present in both cases.

## V. Discussion

It now becomes clear why the lack of velocity profile similarity is important in the figures discussed above. It means that $u_e/u_{ZS}$ is not a constant for the datasets investigated in Figs. 1-3. In fact for the dataset shown in Fig. 3, the ratio $u_e/u_{ZS}$ changes by 300% from smallest to largest value. Besides the three datasets above, we also examined eight additional datasets used in Castillo and George's[1] original paper and that are available in Coles and Hirst.[11] For all of the datasets that we cross-checked, none of the replotted datasets showed similarity in the outer region when plotted as velocity profile data using $u_{ZS}$ scaling. The logical conclusion is that those datasets are showing the false form of defect similarity and do not satisfy all of the velocity profile related similarity criteria developed using the flow governing equation approach[1,13-15] to similarity or the developments given in Appendix A. This then is the nature of the defect profile similarity compliance problem mentioned in the Abstract/Introduction.

The above true and false similarity argument rests on the fact that defect profile similarity must be accompanied by velocity profile similarity. That in turn rests on the condition that $u_e/u_s$ must be a constant for similarity. The first theoretical derivation for this criterion based on the flow governing equation approach to similarity was done by Rotta[13] and Townsend.[14] More recently, Castillo and George[1] and Jones, Nickels, and Marusic[15] also derived the same criterion using flow governing equation approaches. In Appendix A we offer three additional theoretical approaches to prove that this must be the case. The approaches used in Appendix A are all based on the definition of similarity. Therefore there are at least two distinct theoretical routes that have as a similarity requirement $u_e/u_s$=constant. The consequence of Rotta's Eq. 14.3, Townsend's Eq. 7.2.3, Castillo and George's Eq. 9, or Jones, Nickels, and Marusic Eq. 3.9 a2+a4 criterion is not at first evident until one looks at Eqs. 1-2 above or Eq. 6 above. This criterion requires that defect similarity must be accompanied by velocity profile similarity. It is astonishing that there has been no discussion in the literature in this regard. For more than sixty years all discussion in the turbulent boundary layer scaling has been about defect profile similarity when theory indicates that using the traditional velocity profile definition would have worked just as well. That the equivalence was not noticed may also explain why the false similarity premise was not noticed earlier. In any case, results in Appendix A as well as the flow governing equation approaches to similarity by Rotta,[13] Townsend,[14] Castillo and George,[1] and Jones, Nickels, and Marusic[15] require that defect profile similarity must be accompanied by velocity profile similarity.

The consequences of the false similarity premise could be widespread. A quick survey indicates that there are papers in the literature that will need to be re-evaluated in light of the above findings. Any claims of similarity based on the defect profile together with any velocity scaling variable other than $u_e$ need to verify that all criteria for similarity have been satisfied. It is therefore essential to eliminate any possibility that this concept of true and false similarity is somehow incorrect or flawed. In this regard, the most obvious point to examine is whether the Appendix A derivations or the Rotta,[13] Townsend,[14] Castillo and George,[1] and Jones, Nickels, and Marusic[15] derivations could be faulty. Some have argued that the flow governing equation approach to similarity developed by Castillo and George is problematic since it is known[6,7] that $u_e$ makes a poor outer region scaling parameter compared to $u_{ZS}$ or $u_\tau$, the friction velocity. Examination of those two review papers[6,7] shows that the claims are based on experimental data plotted as scaled defect profiles. Could these defect profile plots also be cases of false similarity? Consider the review by Panton[6] in which he claimed that $u_e$ is the wrong outer region scale and that it should be $u_\tau$. To support his claim, in part, he compared plots of some of Österlund's[9] datasets. To test for the possibility of false similarity behavior we plot some of Österlund's[9] data using $u_s = u_\tau$ in Fig. 4. Comparison of Fig. 4a and 4b indicate that the selected profiles do not show true similarity behavior when scaled with $u_\tau$. Panton also claimed that DeGraaff and Eaton's[16] dataset showed similar behavior when scaled with $u_\tau$ whereas $u_e$ did not. In Fig. 5 we plot this dataset using $u_\tau$ as the scaling parameter. Once again the data shows the false similarity symptom. Hence, neither case offered as experimental proof by Panton shows true similarity.

In another review paper Buschmann and Gad-el-Hak[7] also argue that $u_{ZS}$ or $u_\tau$ are better scaling parameters in regards to similar scaling compared to $u_e$. In their paper they also use defect profile plots of some of Österlund's[9] profile data. However, in Figs. 2 and 4 we have already shown that Österlund's datasets do not show true similarity behavior when scaled with $u_{ZS}$ or $u_\tau$. Buschmann and Gad-el-Hak also used some Super Pipe flow data from McKeon, Zagarola, and Smits[17] for comparison. In Fig. 6 we show scaled plots of this dataset using $u_\tau$ as the scaling parameter. This dataset does not display true similarity when scaled with $u_\tau$. Therefore the two review papers[6,7] both made claims that might indicate a problem with the flow governing equation approach to similarity[1,13-15] that can be explained instead using the false similarity premise described above. The point is that if the datasets do not display true similarity then any scaling comparisons are meaningless.

The implications of this false similarity premise could possibly lead to a paradigm shift in the accepted scaling model of turbulent boundary layers. The present paradigm is that velocity profile similarity in turbulent boundary layers is fairly common and that either $u_{ZS}$ or $u_\tau$ are the accepted scaling parameters for the outer region. The scaling parameter $u_e$ is not a valid choice based on experimental comparisons. Therefore the present paradigm necessarily requires that the Appendix A derivations as well as the Rotta, Townsend, Castillo and George, and Jones, Nickels, and Marusic flow governing equation approaches to similarity are all invalid since they require that $u_e$ must be an outer region scaling parameter for similar behavior. On the other hand, we have another possible paradigm by Clauser[2] and Skåre and Krogstad[18] among others who advocate for a boundary layer in which the inner region scaling must change proportionally with the outer region scaling in order to observe similar behavior from station to station. This means that the Rotta[13] velocity ratio $u_e/u_\tau$ criterion must be satisfied since $u_\tau$ is universally accepted as being the inner region scaling parameter. In this paradigm the instances of velocity profile similarity are rare. As Clauser[2] and Skåre and Krogstad[18] demonstrated, it is very difficult to produce turbulent boundary layer similarity in the wind tunnel. Elaborate nonlinear pressure gradients need to be setup in the flow direction to satisfy the Rotta criterion. In this paradigm the Rotta, Townsend, Castillo and George, and Jones, Nickels, and Marusic flow governing equation approaches are valid and correctly predicting similarity in experimental datasets (for example Skåre and Krogstad[18]). So the question of the prevalence of similarity of the turbulent velocity profile will play an important role in determining which paradigm will prevail. As a first step in addressing the prevalence question we choose to concentrate on the Castillo and George's contention that "most" turbulent boundary layers are similarity boundary layers. It is our contention that the results presented herein are supporting the Clauser-Rotta paradigm that most turbulent boundary layer are not similar.

For assurance that the true similarity cases actually do exist we offer Figs. 7-10. The data from Skåre and Krogstad,[18] Smith,[19] Clauser,[2] and Eitel-Amor, Örlü, and Schlatter[20] are plotted using the $\delta_1$ as the length scaling parameter and $u_e$ as the velocity scaling parameter. It can be seen that similar behavior occurs in both the defect profile plot and the velocity profile plot. These four figures demonstrate: 1) that true similarity datasets do exist and 2) that $u_e$ is a legitimate velocity scale for similarity contrary to previous[6,7] pronouncements. It should be noted that Weyburne[12] recently used rigorous mathematics to prove that if similarity exists in a set of boundary layer profiles, then $\delta_1$ must be a similar length scale and $u_e$ must be a similar velocity scale. If one considers the flow governing equation approach result, Appendix A result, the experimental results (Figs. 7-10), and Weyburne's[12] theoretical result, then the scaling parameter $u_e$ is worth further consideration as a possible similarity scaling parameter for the turbulent boundary layer velocity profile.

In spite of the evidence presented above, it is possible that there is an alternative explanation that fits the data equally well that would support the present paradigm. However, for this to happen an explanation is needed to explain how it is that defect profile similarity is present in Figs. 1-6 but velocity profile similarity is not present. At the present time we are not aware of any way to explain the results presented in Figs. 1-6 other than with the true and false similarity explanation given above.

Although we have shown that data comparison claims by Panton[6] and Buschmann and Gad-el-Hak[7] are not valid and have presented experimental evidence (Figs. 7-10) and theoretical evidence[12] that $u_e$ makes a good velocity scaling parameter, there will still be those that insist $u_e$ is not a valid outer region scaling parameter. If $u_e$ is not a valid outer region scaling parameter then the Appendix A derivations and the Rotta, Townsend, Castillo and Jones, Nickels, and Marusic flow governing equation approaches must be invalid. Given that we have offered both theoretical and experimental support for $u_e$ scaling then the burden of proof now lies with those advocating against $u_e$ scaling. This burden of proof needs to be in the form of showing where exactly the flow governing equation approach is breaking down. Furthermore, they need to show how the theoretical routes given Appendix A are flawed.

Note that the same type of flow governing equation approach has been used to successfully develop many of the laminar flow similarity solutions.[8] Both the laminar and turbulent flow governing equation approaches start out by assuming that the *x* and *y* velocity components can be described as a product of a *x*-functional and scaled *y*-functional. In each case these product functionals are then substituted into the combined conservation of mass and the flow-appropriate *x*-momentum balance equation. Similarity requires that each term in the transformed equation must change proportionately as one moves downstream. Equivalently, if one divides through by one of the resulting *x*-groupings from the transformed equation, the requirement now becomes that the *x*-groupings ratios must be constants. The similarity flow conditions like $u_e/u_s$=constant are obtained at this step. So which of these steps is the problem area that invalidates the turbulent flow approach but works for the laminar flow case? Until a convincing argument can be found that invalidates the flow governing equation approach or the theoretical arguments in Appendix A then our interpretation presented herein becomes the *de facto* explanation of the observed results.

## VI. Conclusions

Based on the above analysis, we conclude that claims of similarity based on the defect profile obtained when using any velocity scaling variable other than $u_e$ need to ensure that all criteria for similarity have been satisfied. The majority of Castillo, George, and coworkers' claims for similarity are based on plots of the defect profile using the Zagarola and Smits[5] velocity scaling parameter. None of these claims were checked to ensure $u_e/u_{ZS}$ is a constant as required. Of the nine datasets we have cross checked from Castillo and George's original paper that used $u_{ZS}$, not one of the nine datasets showed similarity of the scaled velocity profile in the outer region thereby indicating noncompliance with the similarity constraints developed in Appendix A and by Rotta,[13] Townsend,[14] Castillo and George,[1] and Jones, Nickels, and Marusic.[15] We cross checked other claims in two boundary layer scaling review papers in the literature[6,7] and found the same problem. Although a more thorough study is called for, we can say with some confidence based on the datasets that have been tested thus far that it is not true that "most" turbulent boundary layers are similarity boundary layers as claimed by Castillo and George.[1]

**Acknowledgement**

The author would like to acknowledge the support of the Air Force Research Laboratory and Gernot Pomrenke at AFOSR. In addition, the author would like to thank the various experimentalists for making their datasets available for inclusion in this manuscript. None of this work would have been possible without their active support.

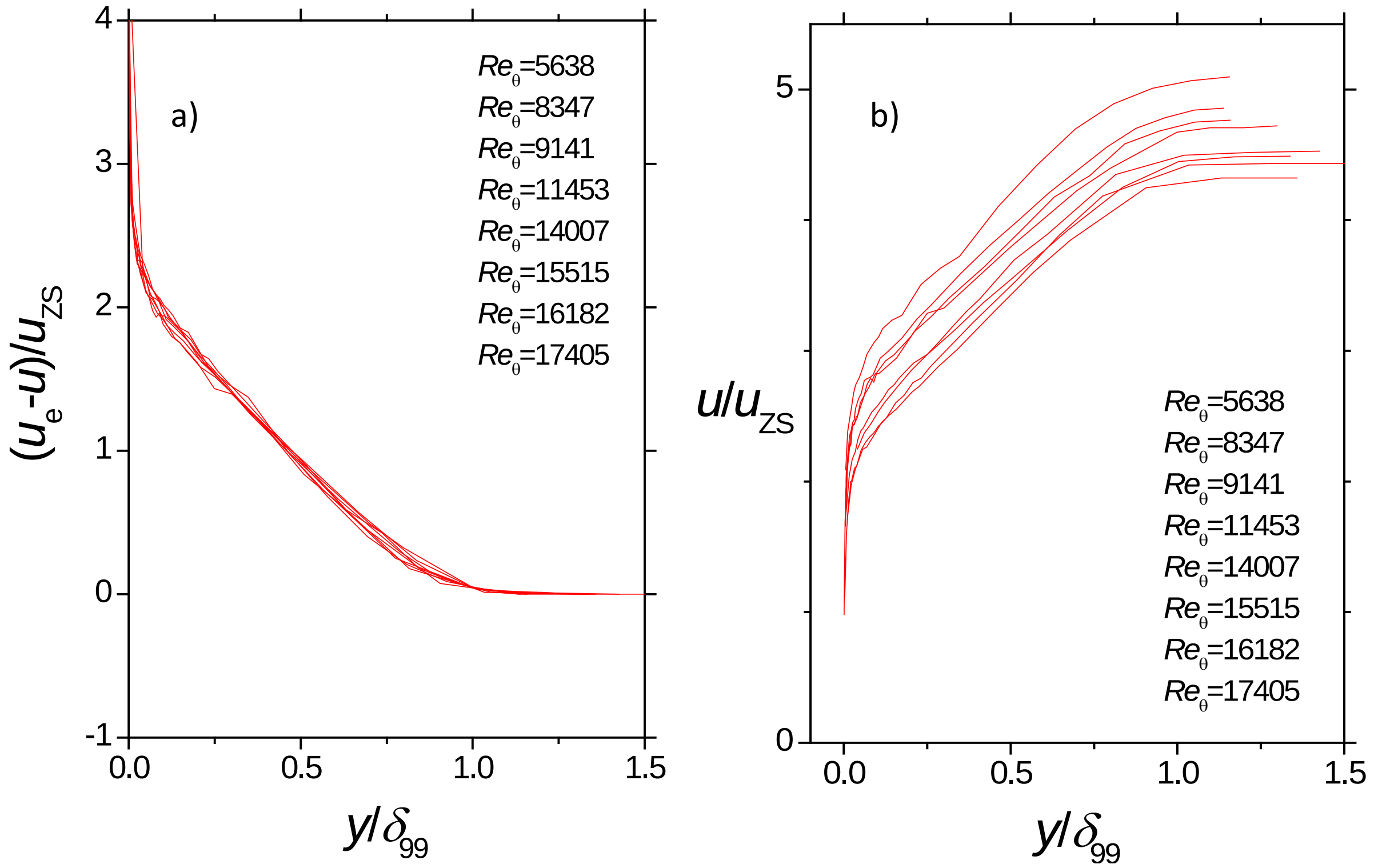

Fig. 1: Clauser's[2] mild APG data plotted as a) defect profiles and b) velocity profiles using $u_s = u_{ZS}$. The $Re_\theta$ list identifies which profiles are plotted.

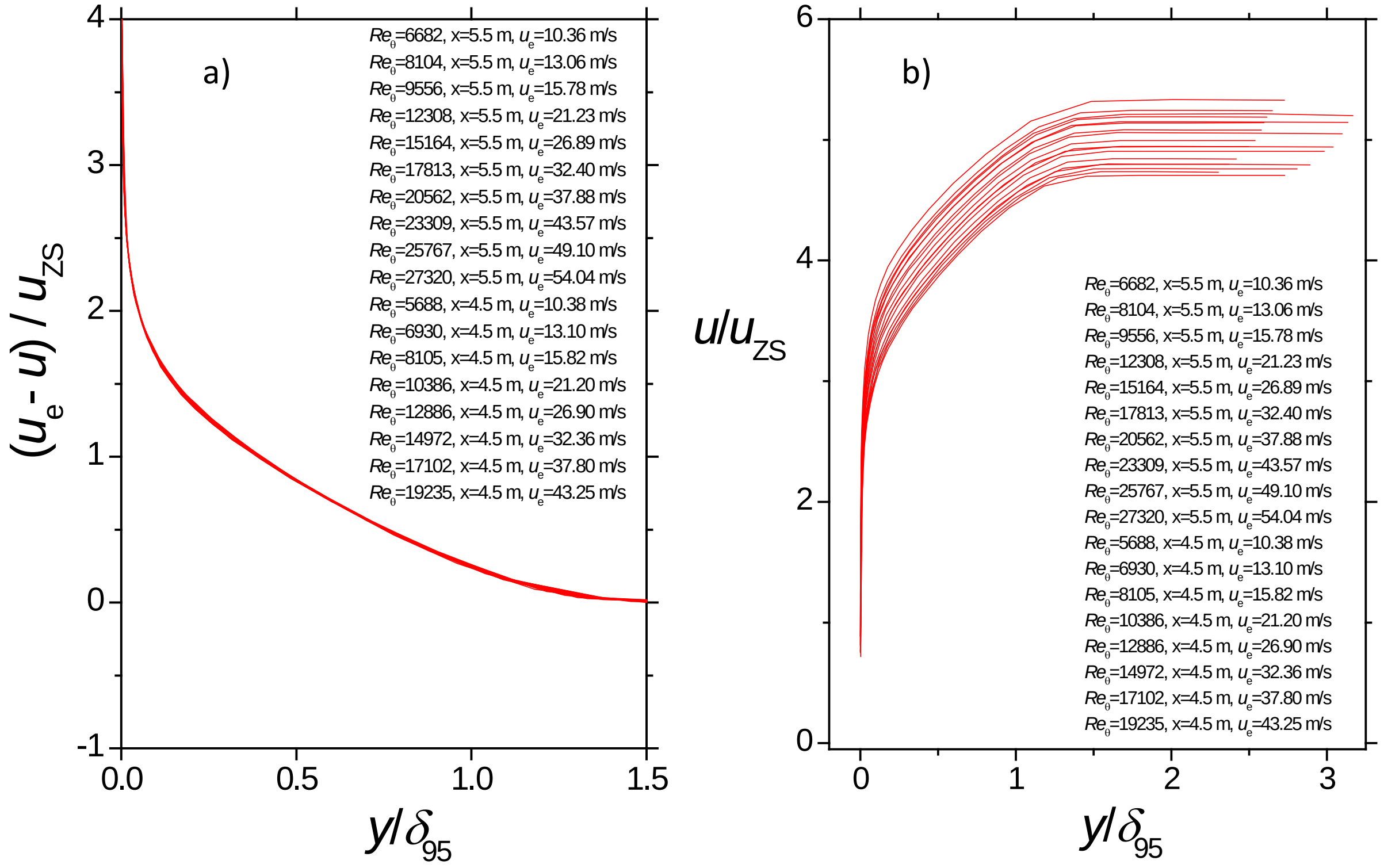

Fig. 2: Österlund's[9] data plotted as a) defect profiles and b) velocity profiles using $u_s = u_{ZS}$. The $Re_\theta$ list identifies which profiles are plotted.

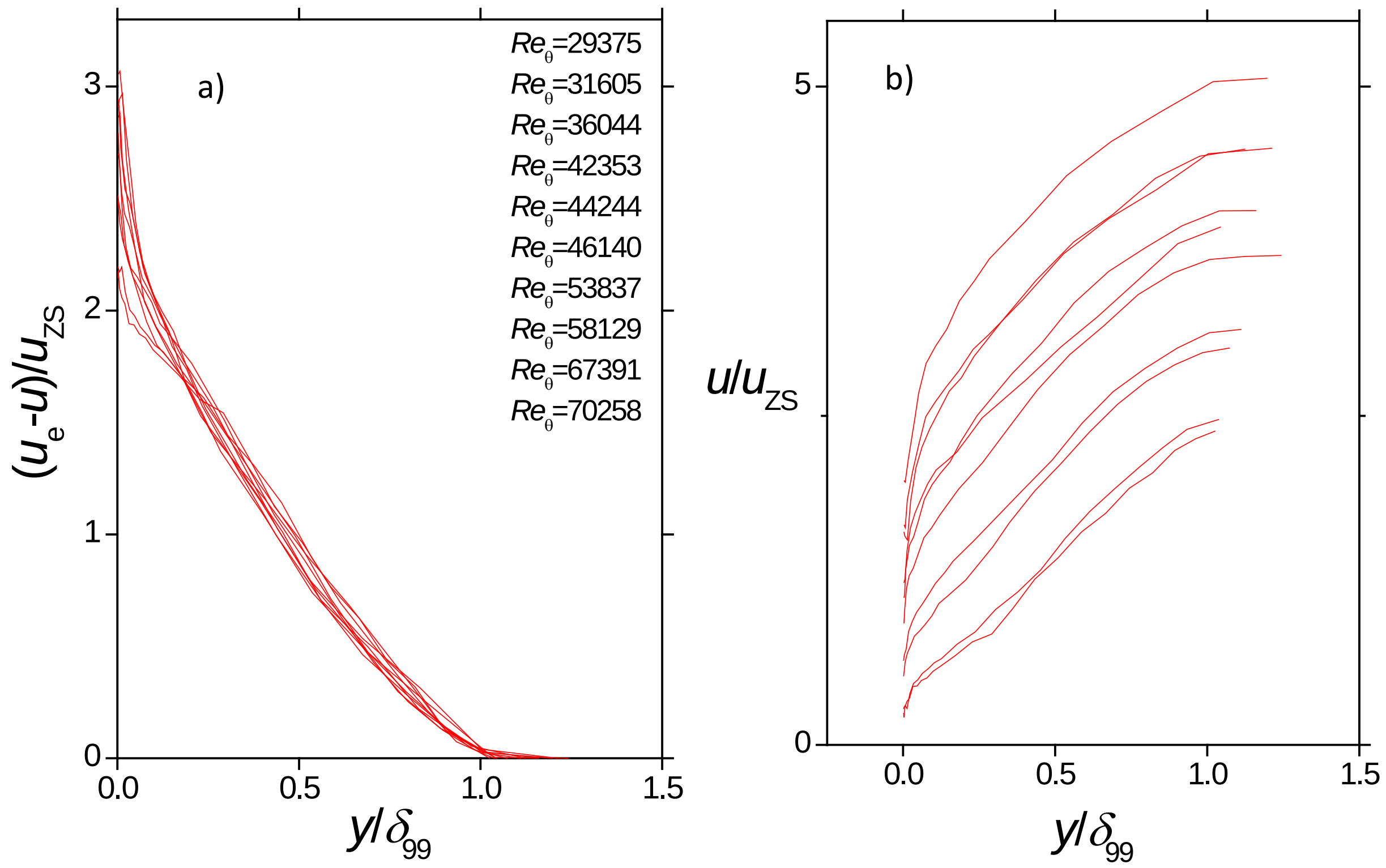

Fig. 3: Schubauer and Klebanoff's[10] data plotted as a) defect profiles and the same data plotted as b) velocity profiles using $u_s = u_{ZS}$. The $Re_\theta$ list identifies which profiles are plotted.

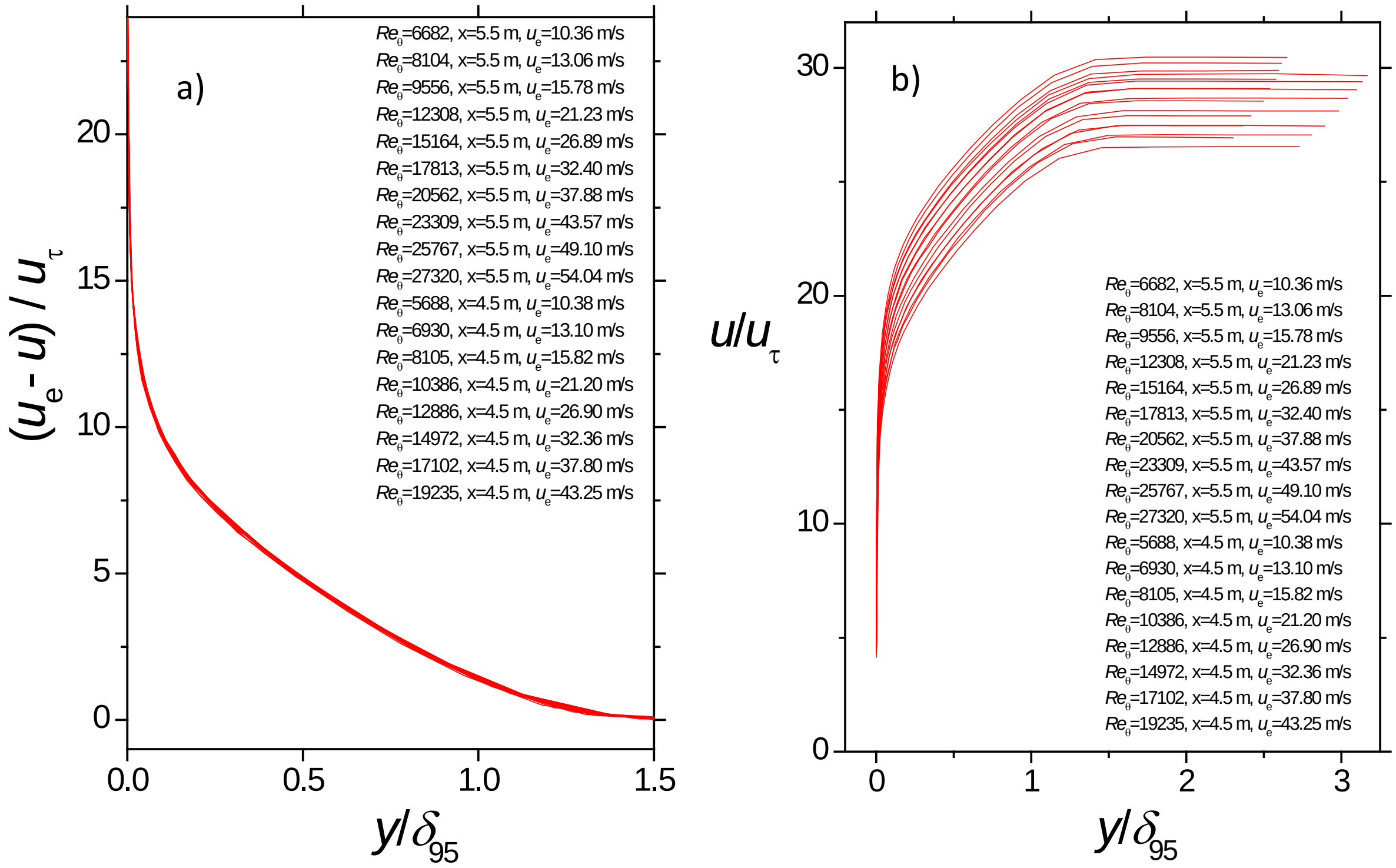

Fig. 4: Österlund's[9] data plotted as a) defect profiles and b) velocity profiles using $u_s = u_\tau$. The $Re_\theta$ list identifies which profiles are plotted.

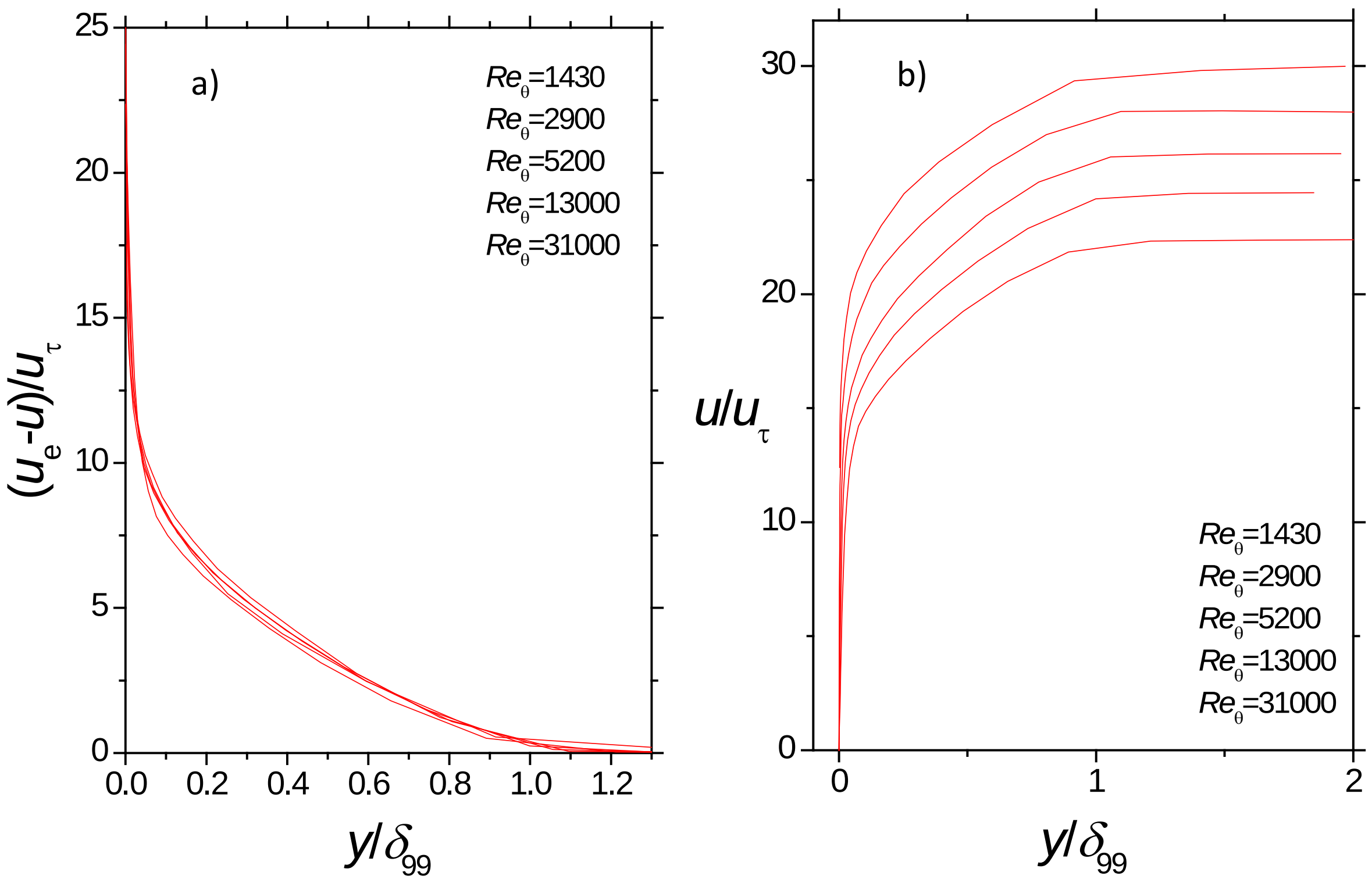

Fig. 5: DeGraaff and Eaton's[16] data plotted as a) defect profiles and same data plotted as b) velocity profiles using $u_s = u_\tau$. The $Re_\theta$ list identifies which profiles are plotted.

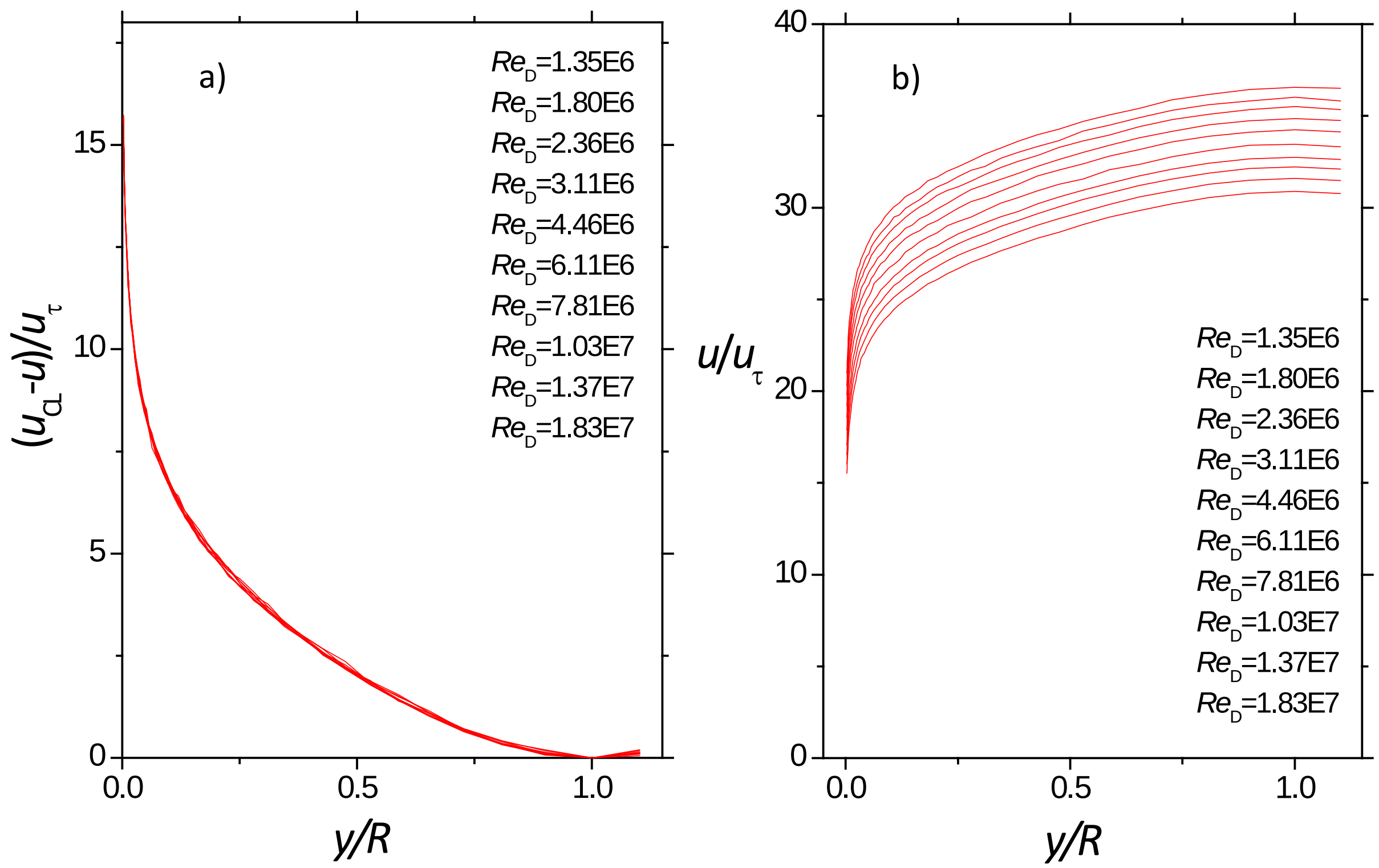

Fig. 6: McKeon, Zagarola, and Smits[17] pipe data plotted as a) scaled defect profiles and b) scaled velocity profiles using $u_s = u_\tau$. The $Re_D$ list identifies which profiles are plotted.

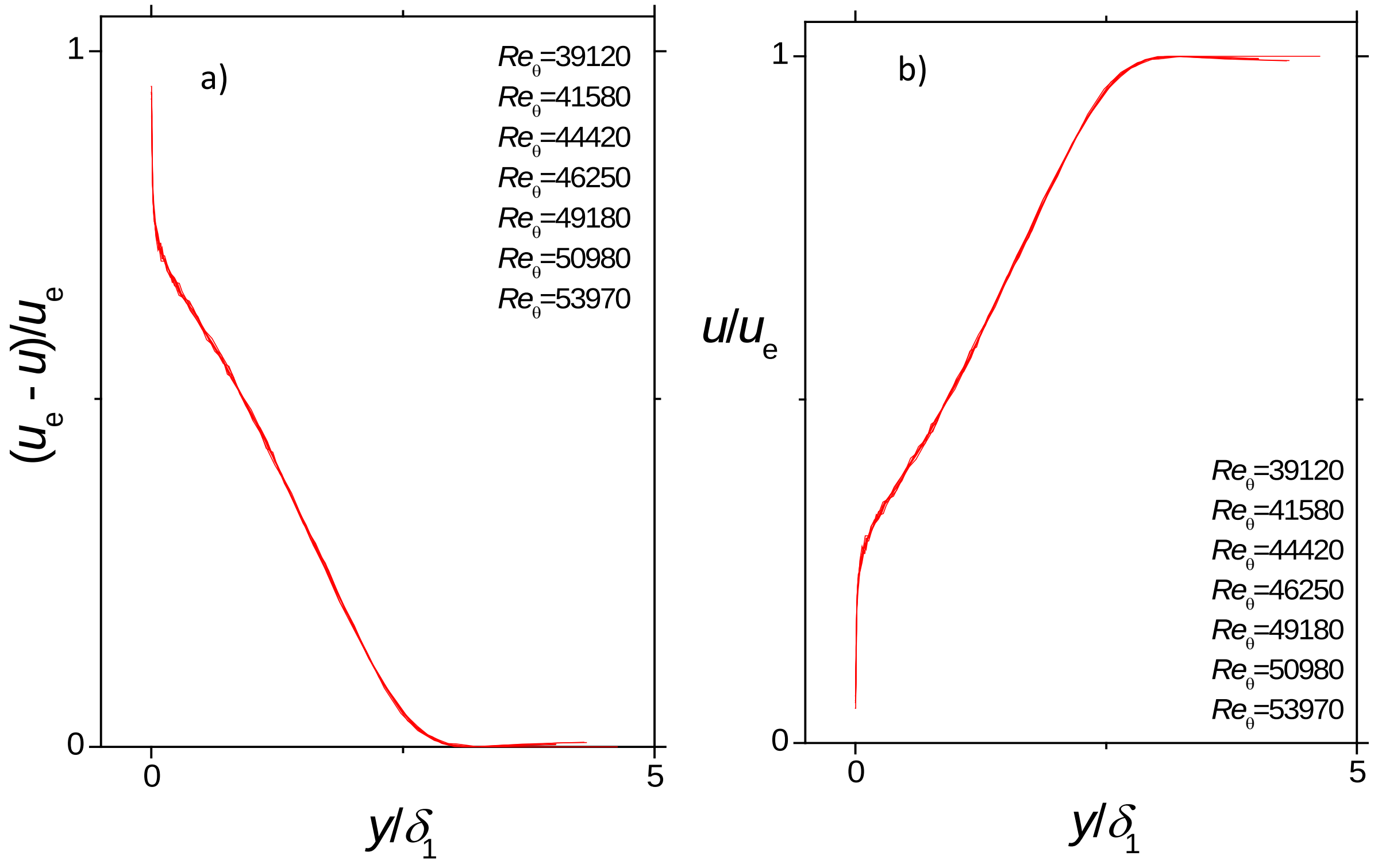

Fig. 7: Skåre and Krogstad's[18] data plotted as a) defect profiles and same data plotted as b) velocity profiles using $u_s = u_e$. The $Re_\theta$ list identifies which profiles are plotted.

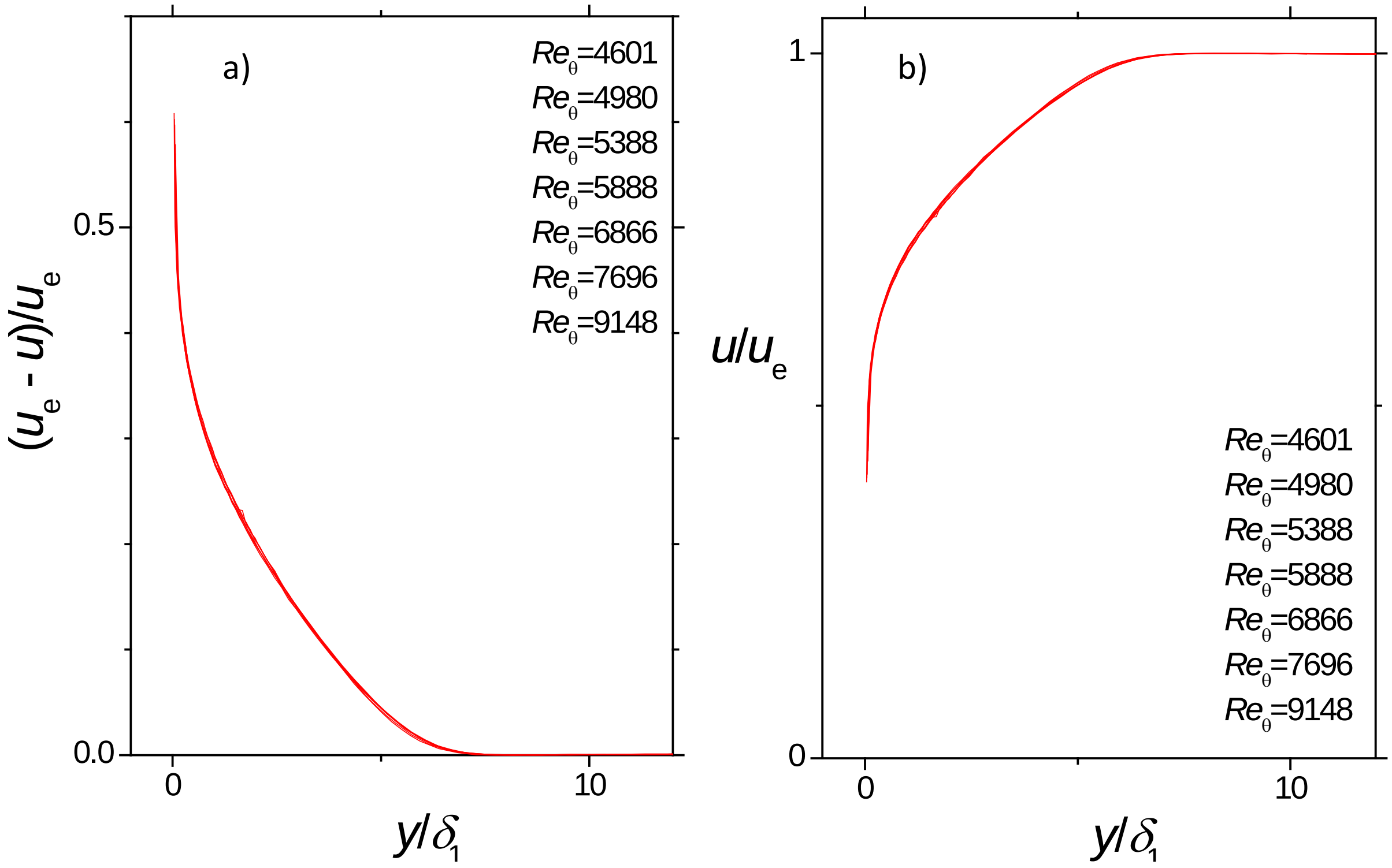

Fig. 8: Smith's[19] data plotted as a) scaled defect profiles and b) scaled velocity profiles using $u_s = u_e$. The $Re_\theta$ list identifies which profiles are plotted.

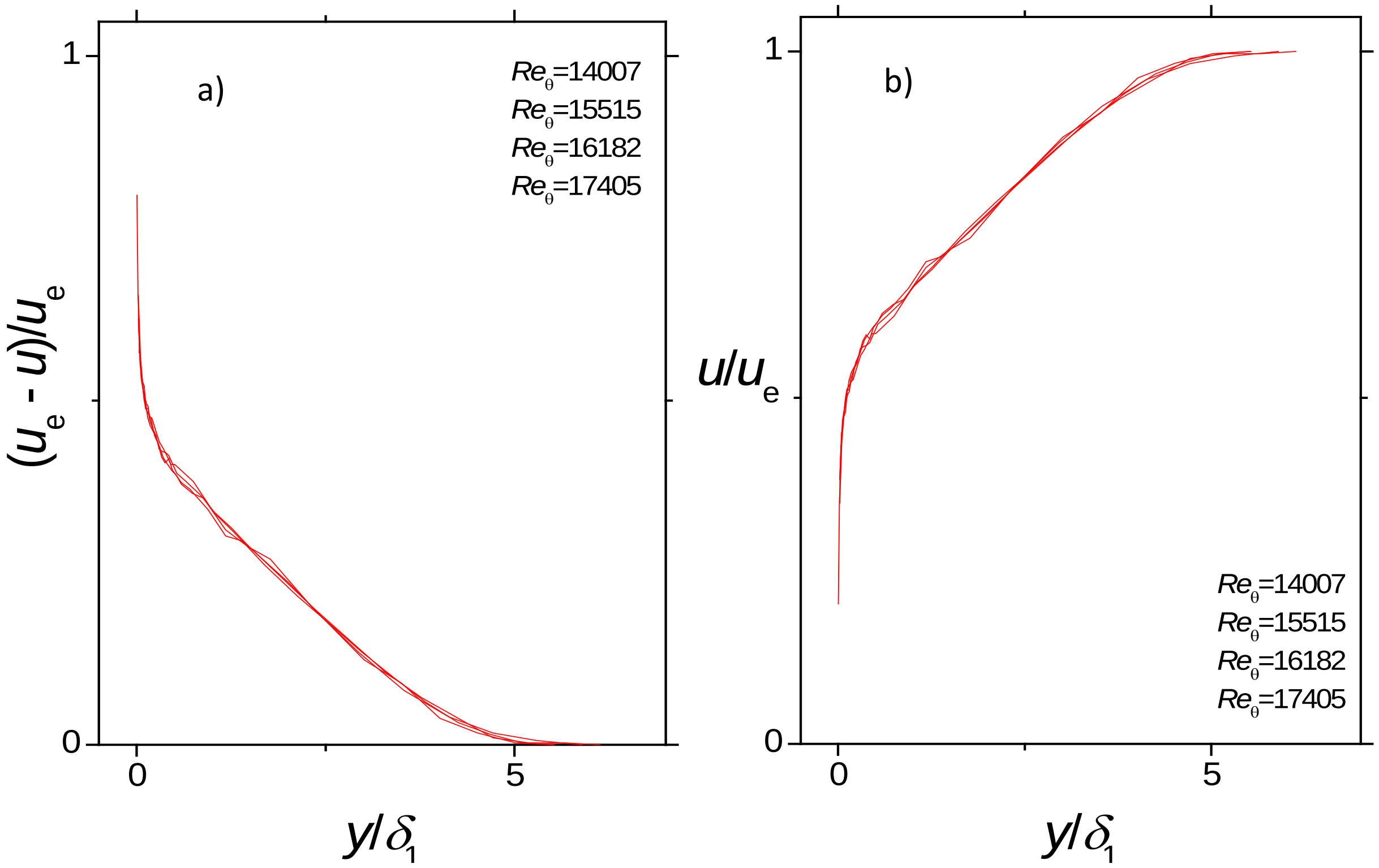

Fig. 9: Clauser's[2] mild APG data plotted as a) defect profiles and same data plotted as b) velocity profiles using $u_s = u_e$. The $Re_\theta$ list identifies which profiles are plotted.

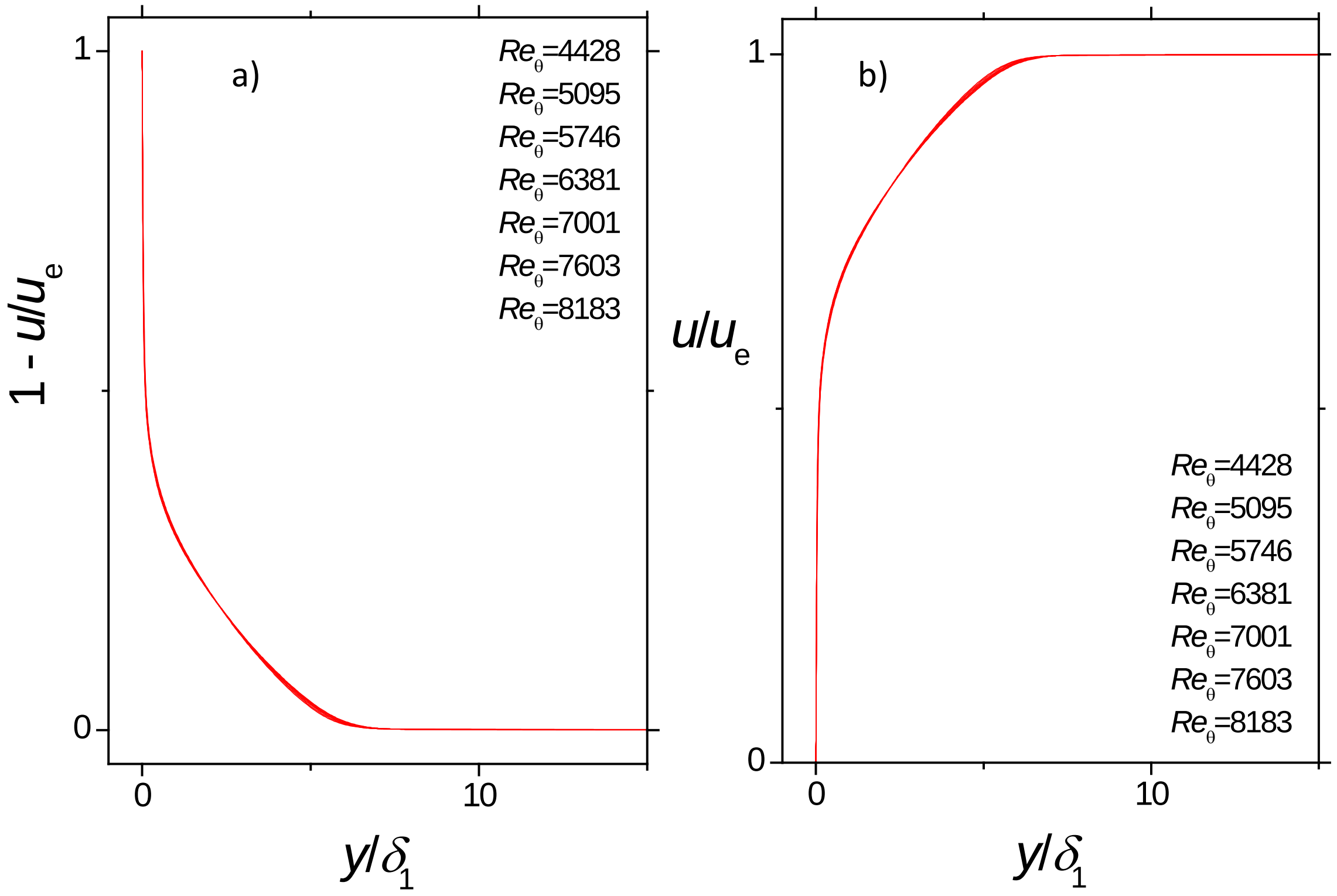

Fig. 10: Eitel-Amor, Örlü, and Schlatter's[20] LES data plotted as a) scaled defect profiles and b) scaled velocity profiles using $u_s = u_e$. The $Re_\theta$ list identifies which profiles are plotted.

## Appendix A: Alternative Derivations of the Velocity Ratio Scaling Criterion

The flow governing approaches to similarity by Rotta[13] (see his Eq. 14.3), Townsend[14] (see his Eq. 7.2.3), and Castillo and George[1] (see their Eq. 9) and Jones, Nickels, and Marusic[15] (see their Eq. 3.9, a2+a4) all have as a similarity requirement that $u_e / u_s$=constant (or equivalent). The flow governing approach is not the only theoretical route to this criterion. We can offer at least three additional simple theoretical routes to substantiate $u_e$ as a valid similarity parameter choice.

First, start with the definition of defect similarity given by

$$\frac{u_e(x) - u(x, y/\delta_s)}{u_s(x)} = f(y/\delta_s) \quad \text{for all y,} \tag{A.1}$$

where *f* is an universal profile function of the dimensionless height. Note that the "for all *y*" part is usually assumed. Since this equation must hold for all *y*; take the limit where $y \to 0$ so that Eq. A.1 becomes

$$\frac{u_e(x)}{u_s(x)} = f(0) \quad , \tag{A.2}$$

where $f(0)$ is some universal constant of similarity. Therefore, $u_e$ must be a valid similarity parameter choice.

Now let us consider a second theoretical route. Weyburne[12] recently used rigorous mathematics to prove that if similarity exists in a set of boundary layer profiles, then the displacement thickness $\delta_1$ must be a similar length scale and $u_e$ must be a similar velocity scale. This derivation starts with the traditional definition of similarity given according to Schlichting[8] as

$$\frac{u(x_1, y/\delta_s(x_1))}{u_s(x_1)} = \frac{u(x_2, y/\delta_s(x_2))}{u_s(x_2)} \quad \text{for all y.} \tag{A.3}$$

If similarity is present in a set of velocity profiles then it is self-evident that the properly scaled first derivative profile curves (derivative with respect to the scaled *y*-coordinate) must also be similar. It is also self-evident that the areas under the scaled first derivative profiles plotted against the scaled *y*-coordinate must be equal for similarity. In mathematical terms, the area under the scaled first derivative profile curve is expressed by

$$a(x) = \int_0^{h/\delta_s} d\left\{\frac{y}{\delta_s}\right\} \frac{d\left\{u(x, y/\delta_s)/u_s\right\}}{d\left\{\dfrac{y}{\delta_s}\right\}} \quad , \tag{A.4}$$

where $a(x)$ is in general a non-zero numerical value and $y = h$ is located deep in the free stream. With a variable switch $\left(d\{y/\delta_s\} \Rightarrow (1/\delta_s)dy\right)$, Eq. A.4 can be shown to reduce to

$$a(x) = \frac{u_e(x)}{u_s(x)} \quad . \tag{A.5}$$

Similarity requires that $a(x_1) = a(x_2) = \text{constant}$. Once again, $u_e$ must be a valid similarity parameter choice.

A third theoretical route also starts with Eq. A.3. Since this equation must hold for all *y*; take the case where $y \to h(x)$ such that $u(x,h(x)) = u_e(x)$. Similarity assumes that we end at the same scaled *y*-value given by

$$\frac{h(x_1)}{\delta_s(x_1)} = \frac{h(x_2)}{\delta_s(x_2)} \quad , \tag{A.6}$$

but this is easily satisfied since we can freely choose $h(x_1)$ and $h(x_2)$ to be located anywhere in the boundary layer edge region. If we choose the *y*-values $h(x_1)$ and $h(x_2)$ to satisfy Eq. A.6 at the boundary layer edge, then Eq. A.3 reduces to

$$\frac{u_e(x_1)}{u_s(x_1)} = \frac{u_e(x_2)}{u_s(x_2)} \quad . \tag{A.7}$$

Hence $u_e$ must be a valid similarity parameter choice.

Therefore we have three additional theoretical derivations, all of which indicate that while other velocity scaling parameters are not excluded, it must be the case that if similarity exists in a set of velocity profiles then the velocity at the boundary layer edge $u_e$ must be a similarity scaling parameter for the 2-D boundary layer. Notice that none of these derivations are based on the flow governing equations but rather on the definition of similarity itself. Hence these arguments apply to any 2-D boundary layer flow that shows similarity from station to station.